\documentclass{article}

\usepackage{jheppub}
\usepackage{tikz}
\usepackage{xcolor}
\usepackage{amsmath}
\usepackage{caption}
\usepackage{setspace}

\begin{document}

\title{Line Operators in the Left-Right Symmetric Model}

\author[a;c]{Jiaming Guo}
\author[a;c]{,\,\,Rui Yang} 
\author[a;b;c;d,1]{and\,\,Xun Xue}

\affiliation[a]{School of Physics, Xinjiang University, Urumqi 830046, China}
\affiliation[b]{School of Physics and Information Engineer, Zhaotong University, Zhaotong 657000, China}
\affiliation[c]{Chongqing Institute of East China Normal University, Chongqing 401120, China}
\affiliation[d]{Department of Physics, East China Normal University, Shanghai 200241, China}

\emailAdd{xxue@phy.ecnu.edu.cn}

\abstract{In this paper, we studied lineoperators in the Left-Right Symmetric Model. The gauge group of  Left-Right Symmetric Model is ${G} = SU(3) \times SU(2)_{L} \times SU(2)_{R} \times U(1)_{B - L}$. We derived the spectrum of line operators in all possible scenarios within left-right symmetric models. We then studied the $\theta$ angles in left-right symmetric model. We also discuss the effect of symmetry breaking on the spectrum of line operators and  $\theta$ angles.}

\maketitle

\section{Introduction}

In the last century, with the proposal of Noether's theorem, the connection between symmetry and conservation laws was revealed. Noether's theorem states that every continuous symmetry corresponds to a conservation law. Noether's theorem reveals the importance of symmetry in physics and provides a method to obtain conserved quantities from symmetries. Subsequently, with the improvement and introduction of Lie group-related theories, theories of symmetry have made significant contributions to the development of quantum field theory and other theories.

In recent years, symmetry theory has been continuously developing, and many new forms of symmetries have been proposed. These symmetries are generally referred to as ``generalized symmetries\cite{cordova2022snowmass}. ''  Well-known generalized symmetries include subsystem symmetries, higher-form symmetries, higher-group symmetries, and non-invertible symmetries  \cite{gomes2023introduction,choi2022noninvertible}.

Ordinary symmetries, due to their transformation parameters being 0-forms, are also called 0-form symmetries. Correspondingly, the conserved current defined by ordinary symmetries is a 1-form. If we extend the parameter of a symmetry to higher form, we define a higher-form symmetry. Because the parameters of ordinary symmetries are 0-forms defined on 0-dimensional regions, they can naturally be related to the transformation of objects also defined on 0-dimensional regions, meaning that the charged objects of ordinary symmetries are local operators. We will consider 1-form symmetry, whose parameter is a 1-form, and the corresponding conserved current is a 2-form. Analogous to the case of ordinary symmetries, the charged objects of 1-form symmetry should be objects defined on 1-dimensional regions, i.e. line operators\cite{gomes2023introduction}.

Line operators, in addition to being the charged objects of 1-form symmetry, have other properties. It is generally believed that the gauge group of the Standard Model is $SU(3)\times SU(2)\times U(1)$. However, since the correlation functions of local operators in $R^{1,3}$ only depend on the Lie algebra of the gauge group, the specific gauge group of the Standard Model has some uncertainty. We can write the accurate gauge group of the Standard Model as follows:
\begin{equation}
G = \frac{\widetilde{G}}{\Gamma} = \frac{SU(3) \times SU(2) \times U(1)}{\Gamma}\ ,
\end{equation}
where
\begin{equation}
\Gamma = Z_{6},Z_{3},Z_{2},\mathbf{1},
\end{equation}
and existing experiments cannot distinguish these possibilities. In \cite{tong2017line},it proposed a possible method to distinguish the specific gauge group of the Standard Model through line operators. The paper mentions that different choice of \(\Gamma\) can affect the spectrum of line operators, the periodicity of the \(\theta\) angle, and the states invariant under time reversal, as well as the minimum values of electric and magnetic charges after electroweak symmetry breaking. In a 2024 paper \cite{alonso2024fractional}, it was proposed that the specific choice of the gauge group of the Standard Model would manifest as different spectrum of fractional-charge particles, which can be experimentally tested through the LHC to distinguish the specific gauge group of the Standard Model.

We are interested in whether methods in \cite{tong2017line} can be generalised to models beyond the Standard Model, so we conducted a similar discussion on the left-right symmetric model. The left-right symmetric model is a theoretical model beyond the Standard Model\cite{maiezza2017higgs,senjanovic2017left}. Compared to the gauge group of the Standard Model, the left-right symmetric model adds a right-handed \(SU(2)_{R}\), i.e. the complete gauge group is \(SU(3) \times SU(2)_{L} \times SU(2)_{R} \times U(1)_{B - L}\). The left-right symmetric model solves the left-right asymmetry problem of the Standard Model by adding the right-handed \(SU(2)_{R}\) and can explain the problem of spontaneous parity breaking\cite{harris2017left}. The left-right symmetry model also introduces new gauge bosons and scalar particles with masses larger than those of the Standard Model particles, providing more possibilities for experimental detection. Due to the addition of the right-handed \(SU(2)_{R}\), the electroweak symmetry breaking process of the left-right symmetric model has some differences from that of the Standard Model, i.e., compared to the \(SU(2) \times U(1)_{Y} \rightarrow U(1)_{em}\) of the Standard Model, the left-right symmetric model breaks to the electromagnetic theory in two steps, i.e. \(SU(2)_{L} \times SU(2)_{R} \times U(1)_{B - L} \rightarrow SU(2)_{L} \times U(1)_{Y} \rightarrow U(1)_{em}\)\cite{maiezza2017higgs}.

The content of this paper is organized as follows. Chapter 2 discusses the effect of different gauge groups of the left-right symmetric model on the spectrum of line operators. Chapter 3 discusses the effect of different gauge groups on the periodicity of the \(\theta\) angle and the \(\theta\) angle invariant under time reversal. Chapter 4 discusses the changes in the spectrum of line operators during the electroweak symmetry breaking process.

\section{Spectrum of line operators}

The gauge group of the minimal left-right symmetric model is:
\begin{equation}
\widetilde{G} = SU(3) \times SU(2)_{L} \times SU(2)_{R} \times U(1)_{B - L},
\end{equation}
with two \(SU(2)\) groups exchanging symmetry. Here, \(SU(3)\) and \(SU(2)_{L}\) correspond to the \(SU(3)\) and \(SU(2)\) of the Standard Model, respectively. The \(B\) and \(L\) in \(U(1)_{B - L}\) represent baryon number and lepton number, respectively. Just as the gauge group of the Standard Model can actually be taken as:

\begin{equation}
G = \frac{SU(3) \times SU(2) \times U(1)_{Y}}{\Gamma},
\end{equation}
where \(\Gamma = 1,Z_{2},Z_{3},Z_{6}\),the gauge group of the left-right symmetric model can also be taken as:
\begin{equation}
G = \frac{\widetilde{G}}{\Gamma} = \frac{SU(3) \times SU(2)_{L} \times SU(2)_{R} \times U(1)_{B - L}}{\Gamma},
\end{equation}
where \(\Gamma = 1,Z_{2L},Z_{2R},Z_{2L} \times Z_{2R},Z_{3},Z_{2L} \times Z_{3},Z_{2R} \times Z_{3},Z_{2L} \times Z_{2R} \times Z_{3}\). Since \(\Gamma = Z_{2L} \times Z_{2R}\) does not have a unique generator,~different combinations of the two \(SU(2)\) groups with other gauge groups will lead to different spectrum of line operators. We will list the spectrum of line operators under different combinations of gauge groups later.

\par
Since the left-right symmetric model has four gauge groups, the line operators are labeled by four electric charges and four magnetic charges, each pair of electric and magnetic charges corresponding to a gauge field, i.e. \(z_{3}^{e},z_{3}^{m} = 0,1,2\) for the electric and magnetic charges of \(SU(3)\), \(z_{2L}^{e},z_{2L}^{m} = 0,1\) for the electric and magnetic charges of \(SU(2)_{L}\), \(z_{2R}^{e},z_{2R}^{m} = 0,1\) for the electric and magnetic charges of \(SU(2)_{R}\). For \(U(1)_{B - L}\), since \(B - L\) is at least \(\frac{1}{3}\), for convenience, we can take \(\frac{q}{3} = B - L\), \(q \in Z\) be the electric charge of \(U(1)_{B - L}\), and \(g\) be the magnetic charge of \(U(1)_{B - L}\), and we choose normalization such that the magnetic charge \(g\) takes integer values when there is no quotient group.

\par
Considering the Dirac quantization condition between pure electric and pure magnetic line operators, we get:
\begin{equation}
e^{\frac{2}{3}\pi iz_{3}^{e}z_{3}^{m}}e^{\frac{2}{2}\pi iz_{2L}^{e}z_{2L}^{m}}e^{\frac{2}{2}\pi iz_{2R}^{e}z_{2R}^{m}}e^{- 2\pi iqg} = 1.
\end{equation}
The above Dirac quantization condition can also be written as:
\begin{equation}
2z_{3}^{e}z_{3}^{m} + 3z_{2L}^{e}z_{2L}^{m} + 3z_{2R}^{e}z_{2R}^{m} - 6qg \in 6Z.
\end{equation}

\par
Next, we will discuss the spectrum of line operators with \(\theta\) angles equal to 0 under different combinations of gauge groups and different values of \(\Gamma\). We will discuss the effect of the \(\theta\) angles on the spectrum of line operators later.

(1).\(\widetilde{G} = (SU(3) \times SU(2)_{L} \times U(1)_{B - L}) \times SU(2)_{R}\)

In this gauge group combination, the generators of \(\Gamma = Z_{2L} \times Z_{2R} \times Z_{3}\) can be written as follows:
\begin{equation}
\xi = e^{\frac{2}{3}\pi iz_{3}^{e}} \otimes e^{\frac{2}{2}\pi iz_{2L}^{e}} \otimes e^{\frac{2}{3}\pi iq},
\end{equation}
\begin{equation}
\chi = e^{\frac{2}{2}\pi iz_{2R}^{e}},
\end{equation}
\begin{equation}
\xi \otimes \chi.
\end{equation}

\par
When \(\Gamma = 1\), the Wilson lines are invariant under the action of \(\xi^{6} \otimes \chi^{2}\), so the electric charges of the Wilson lines can be \(z_{2L}^{e},z_{2R}^{e} = 0,1;z_{3}^{e} = 0,1,2;q \in Z\). According to the Dirac quantization condition, the magnetic charges of the t'Hooft lines can be \(z_{3}^{m},z_{2L}^{m},z_{2R}^{m} = 0;g \in Z\).

\par
When \(\Gamma = Z_{2L}\), the Wilson lines are invariant under the action of \(\xi^{3} \otimes \chi^{2}\), which requires \(\frac{3}{2}z_{2L}^{e} \in Z\), i.e. \(z_{2L}^{e} = 0\). Therefore, the electric charges of the Wilson lines can be \(z_{2L}^{e} = 0;z_{2R}^{e} = 0,1;z_{3}^{e} = 0,1,2;q \in Z\). According to the Dirac quantization condition, the magnetic charges of the t'Hooft lines can be \(z_{2R}^{m},z_{3}^{m} = 0;g \in Z\), and since \(z_{2L}^{e} = 0\), the corresponding \(z_{2L}^{m}\) can be 0,1.

\par
When \(\Gamma = Z_{2R}\), the Wilson lines are invariant under the action of \(\xi^{6} \otimes \chi^{1}\), which requires \(\frac{1}{2}z_{2R}^{e} \in Z\), i.e. \(z_{2R}^{e} = 0\). Therefore, the electric charges of the Wilson lines can be \(z_{2L}^{e} = 0,1;z_{2R}^{e} = 0;z_{3}^{e} = 0,1,2;q \in Z\). According to the Dirac quantization condition, the magnetic charges of the t'Hooft lines can be \(z_{2L}^{m},z_{3}^{m} = 0;g \in Z\), and since \(z_{2R}^{e} = 0\), the corresponding \(z_{2R}^{m}\) can be 0,1.

\par
When \(\Gamma = Z_{2L} \times Z_{2R}\), the Wilson lines are invariant under the action of \(\xi^{3} \otimes \chi^{1}\), which requires \(z_{2L}^{e},z_{2R}^{e} = 0\). Therefore, the electric charges of the Wilson lines can be \(z_{2L}^{e},z_{2R}^{e} = 0;z_{3}^{e} = 0,1,2;q \in Z\). According to the Dirac quantization condition, the magnetic charges of the t'Hooft lines can be \(z_{2L}^{m},z_{2R}^{m} = 0,1;z_{3}^{m} = 0;g \in Z\).

\par
When \(\Gamma = Z_{3}\), the Wilson lines are invariant under the action of \(\xi^{2} \otimes \chi^{2}\), so \(\frac{2}{3}\left( z_{3}^{e} + q \right) = n\), where \(n\) is an integer, i.e. \(q = 3n - z_{3}^{e}\). Therefore, the electric charges of the Wilson lines can be \(z_{2L}^{e},z_{2R}^{e} = 0,1;z_{3}^{e} = 0,1,2;q = 3n - z_{3}^{e}\). According to the Dirac quantization condition, the magnetic charges of the t'Hooft lines can be \(z_{2L}^{m},z_{2R}^{m} = 0;z_{3}^{m} = 0,1,2;g = m - \frac{1}{3}z_{3}^{m}\), where \(m\) is an integer.

\par
When \(\Gamma = Z_{2L} \times Z_{3}\) the Wilson lines are invariant under the action of \(\xi^{1} \otimes \chi^{2}\), so \(\frac{1}{3}z_{3}^{e} + \frac{1}{2}z_{2L}^{e} + \frac{1}{3}q = n\), where \(n\) is an integer, i.e. \(z_{2L}^{e} = 0;q = \begin{pmatrix} 3n - z_{3}^{e} \\ \end{pmatrix}\). Therefore, the electric charges of the Wilson lines can be \(z_{2L}^{e} = 0;z_{2R}^{e} = 0,1;z_{3}^{e} = 0,1,2;q = \begin{pmatrix} 3n - z_{3}^{e} \\ \end{pmatrix}\). According to the Dirac quantization condition, the magnetic charges of the t'Hooft lines can be \(z_{2L}^{m} = 0,1;z_{2R}^{m} = 0;z_{3}^{m} = 0,1,2;g = m - \frac{1}{3}z_{3}^{m}\), where \(m\) is an integer.

\par
When \(\Gamma = Z_{2R} \times Z_{3}\) the Wilson lines are invariant under the action of \(\xi^{2} \otimes \chi^{1}\), so \(\frac{2}{3}\left( q + z_{3}^{e} \right) = n\), i.e. \(q = 3n - z_{3}^{e}\). Therefore, the electric charges of the Wilson lines can be \(z_{2L}^{e} = 0,1;z_{2R}^{e} = 0;z_{3}^{e} = 0,1,2;q = 3n - z_{3}^{e}\). According to the Dirac quantization condition, the magnetic charges of the t'Hooft lines can be \(z_{2L}^{m} = 0;z_{2R}^{m} = 0,1;z_{3}^{m} = 0,1,2;g = m - \frac{1}{3}z_{3}^{m}\).

\par
When \(\Gamma = Z_{2L} \times Z_{2R} \times Z_{3}\) the Wilson lines are invariant under the action of \(\xi^{1} \otimes \chi^{1}\), so \(\frac{1}{3}z_{3}^{e} + \frac{1}{2}z_{2L}^{e} + \frac{1}{3}q = n\), i.e. \(z_{2L}^{e} = 0;q = \begin{pmatrix} 3n - z_{3}^{e} \\ \end{pmatrix}\). Therefore, the electric charges of the Wilson lines can be \(z_{2L}^{e} = 0,1;z_{2R}^{e} = 0;z_{3}^{e} = 0,1,2;q = 3n - z_{3}^{e}\). According to the Dirac quantization condition, the magnetic charges of the t'Hooft lines can be \(z_{2L}^{m} = 0;z_{2R}^{m} = 0,1;z_{3}^{m} = 0,1,2;g = m - \frac{1}{3}z_{3}^{m}\).

(2).\(\widetilde{G} = (SU(3) \times SU(2)_{L}) \times \left( SU(2)_{R} \times U(1)_{B - L} \right)\)

\par
In this gauge group combination, the generators of
\(\Gamma = Z_{2L} \times Z_{2R} \times Z_{3}\) can be written as
follows:
\begin{equation}
\xi = e^{\frac{2}{3}\pi iz_{3}^{e}} \otimes e^{\frac{2}{2}\pi iz_{2L}^{e}},
\end{equation}
\begin{equation}
\chi = e^{\frac{2}{2}\pi iz_{2R}^{e}} \otimes e^{\frac{2}{3}\pi iq},
\end{equation}
\begin{equation}
\xi \otimes \chi.
\end{equation}

\par
When \(\Gamma = 1\), the Wilson lines are invariant under the action of \(\xi^{6} \otimes \chi^{2}\),so \(\frac{2}{3}q = n\), i.e. \(q = 3n\). So the electric charges of the Wilson lines can be \(z_{2L}^{e},z_{2R}^{e} = 0,1;z_{3}^{e} = 0,1,2;q = 3n\).According to the Dirac quantization condition, the magnetic charges of the t'Hooft lines can be \(z_{2L}^{m},z_{2R}^{m},z_{3}^{m} = 0;g \in Z\).

\par
When \(\Gamma = Z_{2L}\), the Wilson lines are invariant under the action of \(\xi^{3} \otimes \chi^{2}\), which requires \(\frac{3}{2}z_{2L}^{e} = 0;\frac{2}{3}q = n\), i.e. \(q = 3n;z_{2L}^{e} = 0\). Therefore, the electric charges of the Wilson lines can be \(z_{2L}^{e} = 0;z_{2R}^{e} = 0,1;z_{3}^{e} = 0,1,2;q = 3n\). According to the Dirac quantization condition, the magnetic charges of the t'Hooft lines can be \(z_{2L}^{m} = 0,1;z_{2R}^{m},z_{3}^{m} = 0;g \in Z\).

\par
When \(\Gamma = Z_{2R}\), the Wilson linse are invariant under the action of \(\xi^{6} \otimes \chi^{1}\), which requires \(\frac{1}{3}q + \frac{1}{2}z_{2R}^{e} = n\), i.e. \(z_{2R}^{e} = 0;q = 3n\). Therefore, the electric charges of the Wilson lines can be \(z_{2L}^{e} = 0,1;z_{2R}^{e} = 0;z_{3}^{e} = 0,1,2;q = 3n\). According to the Dirac quantization condition, the magnetic charges of the t'Hooft lines can be \(z_{2R}^{m} = 0,1;z_{2L}^{m},z_{3}^{m} = 0;g \in Z\).

\par
When \(\Gamma = Z_{2L} \times Z_{2R}\), the Wilson lines are invariant under the action of \(\xi^{3} \otimes \chi^{1}\), which requires \(\frac{3}{2}z_{2L}^{e} = 0;\frac{1}{2}z_{2R}^{e} + \frac{1}{3}q = n\). Therefore, the electric charges of the Wilson lines can be \(z_{2L}^{e} = 0;z_{2R}^{e} = 0;z_{3}^{e} = 0,1,2;q = 3n\). According to the Dirac quantization condition, the magnetic charges of the t'Hooft lines can be \(z_{2L}^{m},z_{2R}^{m} = 0,1;z_{3}^{m} = 0;g \in Z\).

\par
When \(\Gamma = Z_{3}\), the Wilson lines are invariant under the action of \(\xi^{2} \otimes \chi^{2}\), so \(\frac{2}{3}z_{3}^{e} = n;\frac{2}{3}q = n\), i.e. \(z_{3}^{e} = 0;q = 3n\). Therefore, the electric charges of the Wilson lines can be \(z_{2L}^{e},z_{2R}^{e} = 0,1;z_{3}^{e} = 0;q = 3n\). According to the Dirac quantization condition, the magnetic charges of the t'Hooft lines can be \(z_{2L}^{m},z_{2R}^{m} = 0;z_{3}^{m} = 0,1,2;g \in Z\).

\par
When \(\Gamma = Z_{2L} \times Z_{3}\) the Wilson lines are invariant under the action of \(\xi^{1} \otimes \chi^{2}\), so \(\frac{1}{3}z_{3}^{e} = \frac{1}{2}z_{2L}^{e} = 0;\frac{2}{3}q = n\), i.e. \(q = 3n;z_{3}^{e} = z_{2L}^{e} = 0\). Therefore, the electric charges of the Wilson lines can be \(z_{2L}^{e} = 0;z_{2R}^{e} = 0,1;z_{3}^{e} = 0;q = 3n\). According to the Dirac quantization condition, the magnetic charges of the t'Hooft lines can be \(z_{2L}^{m} = 0,1;z_{2R}^{m} = 0;z_{3}^{m} = 0,1,2;g \in Z\).

\par
When \(\Gamma = Z_{2R} \times Z_{3}\) the Wilson lines are invariant under the action of \(\xi^{2} \otimes \chi^{1}\), so \(\frac{2}{3}z_{3}^{e} = n;\frac{1}{3}q + \frac{1}{2}z_{2R}^{e} = n\), i.e. \(z_{3}^{e} = 0;z_{2R}^{e} = 0;q = 3n\). Therefore, the electric charges of the Wilson lines can be \(z_{2L}^{e} = 0,1;z_{2R}^{e} = 0;z_{3}^{e} = 0;q = 3n\). According to the Dirac quantization condition, the magnetic charges of the t'Hooft lines can be \(z_{2L}^{m} = 0;z_{2R}^{m} = 0,1;z_{3}^{m} = 0,1,2;g \in Z\).

\par
When \(\Gamma = Z_{2L} \times Z_{2R} \times Z_{3}\) the Wilson lines are invariant under the action of \(\xi^{1} \otimes \chi^{1}\), so \(\frac{1}{3}z_{3}^{e} + \frac{1}{2}z_{2L}^{e} = n;\frac{1}{2}z_{2R}^{e} + \frac{1}{3}q = 1\). Therefore, the electric charges of the Wilson lines can be \(z_{2L}^{e} = 0;z_{2R}^{e} = 0;z_{3}^{e} = 0;q = 3n\). According to the Dirac quantization condition, the magnetic charges of the t'Hooft lines can be \(z_{2L}^{m},z_{2R}^{m} = 0,1;z_{3}^{m} = 0,1,2;g \in Z\).

In $\widetilde{G} = \left( SU(3) \times U(1)_{B - L} \right) \times SU(2)_{L} \times SU(2)_{R} $ , the spectrum of line operators is the same as the (1) case. In $\widetilde{G} = SU(3) \times \left( SU(2)_{L} \times U(1)_{B - L} \right) \times SU(2)_{R}, SU(3) \times SU(2)_{L} \times SU(2)_{R} \times U(1)_{B - L}$ and $(SU(3) \times SU(2)_{L}) \times \left( SU(2)_{R} \right) \times (U(1)_{B - L}) $ cases, the spectrum of line operators is the same as the (2) case.

\par
Due to the symmetry between  $SU(2)_{L}$ and  $SU(2)_{R}$, the above results omit cases that can be obtained by exchanging \(SU(2)_{L}\) and \(SU(2)_{R}\).

\section{$\theta$ angles of L-R symmetric model}

In this section, we will discuss the \(\theta\) angles in the left-right symmetric model. In general, when $\theta \neq 0$, the $\theta$ angle changes the spectrum of line operators through  Witten effect. For example, for $SU(N)$, when $\theta = 2\pi$, its electric charge $z^e$ of the line operator will pick up a $z^m$ and becomes $z^e + z^m$. Similar to the Standard Model case, each part of the gauge field corresponds to a \(\theta\) angle. Corresponding to \(SU(2)_{L},SU(2)_{R},SU(3)\) and \(U(1)_{B - L}\), we refer to each \(\theta\) angle as \(\theta_{2L},\theta_{2R},\theta_{3}\) and \(\widetilde{\theta}\) respectively. Taking quotient group by the center of the simple gauge group will expand the period of the \(\theta\) angle. For example, in the following case:
\begin{equation}
U(N) = \frac{SU(N) \times U(1)}{Z_{N}},
\end{equation}
where \(N = 2,3\). The gauge field and its field strength corresponding to the \(U(1)\) gauge group are denoted as \(\widetilde{a}\) and \(\widetilde{f}\), and the gauge field and field strength corresponding to the \(SU(N)\) gauge group are denoted as \(a\) and \(f\). The \(\theta\) term of \(SU(N) \times U(1)\) can be written as:
\begin{equation}
S_{\theta} = \frac{\theta_{N}}{16\pi^{2}}\int_{}^{}{tr(f \star f)} + \frac{\widetilde{\theta}}{16\pi^{2}}\int_{}^{}\widetilde{f} \star \widetilde{f}.
\end{equation}
After taking the quotient, the corresponding gauge field of the \(U(N)\) group can be written as:
\begin{equation}
b = a + \widetilde{a}\mathbf{1}_{N}.
\end{equation}
The corresponding field strength, denoted as \(g\), can also be expressed in terms of the field strengths of \(SU(N)\) and \(U(1)\) as:
\begin{equation}
g = f + \widetilde{f}\mathbf{1}_{N}.
\end{equation}
Thus, the theta term of \(SU(N) \times U(1)\) can be rewritten in terms of \(b\) and \(g\) as:
\begin{equation}
\begin{split}
S_{\theta} &= \frac{\theta_{N}}{16\pi^{2}}\int_{}^{}{tr(f \star f)} + \frac{\widetilde{\theta}}{16\pi^{2}}\int_{}^{}\widetilde{f} \star \widetilde{f} \\
&= \frac{\theta_{N}}{16\pi^{2}}\int_{}^{}{tr(g \star g)} + \frac{\widetilde{\theta} - \theta_{N}N}{16\pi^{2}N^{2}}\int_{}^{}{\left\lbrack tr(g) \right\rbrack \star \left\lbrack tr(g) \right\rbrack}.
\end{split}
\end{equation}
It can be seen that the ranges of \(\theta_{N}\) and \(\widetilde{\theta}\) are \(\theta_{N} \in \lbrack 0,2\pi)\) and \(\widetilde{\theta} \in \lbrack 0,2\pi N^{2})\).

For \(\frac{SU(2) \times SU(3) \times U(1)}{Z_{2} \times Z_{3}}\), the
theta term in this case is of the following form:
\begin{equation}
S_{\theta} = \frac{3\theta_{2}}{16\pi^{2}}\int_{}^{}{tr(f_{2} \otimes \mathbf{1}_{3} \star f_{2} \otimes \mathbf{1}_{3})} + \frac{2\theta_{3}}{16\pi^{2}}\int_{}^{}{tr(f_{3} \otimes \mathbf{1}_{2} \star f_{3} \otimes \mathbf{1}_{2})} + \frac{\widetilde{\theta}}{16\pi^{2}}\int_{}^{}\widetilde{f} \star \widetilde{f}.
\end{equation}
We can define the following new gauge fields and corresponding field strengths:
\begin{equation}
b_{2} = a_{2} + \widetilde{a}\mathbf{1}_{2},g_{2} = f_{2} + \widetilde{f}\mathbf{1}_{2},
\end{equation}
\begin{equation}
b_{3} = a_{3} + \widetilde{a}\mathbf{1}_{3},g_{3} = f_{3} + \widetilde{f}\mathbf{1}_{3},
\end{equation}
\begin{equation}
\begin{split}
S_{\theta} &= \frac{3\theta_{2}}{16\pi^{2}}\int_{}^{}{tr(f_{2} \otimes \mathbf{1}_{3} \star f_{2} \otimes \mathbf{1}_{3})} + \frac{2\theta_{3}}{16\pi^{2}}\int_{}^{}{tr(f_{3} \otimes \mathbf{1}_{2} \star f_{3} \otimes \mathbf{1}_{2})} + \frac{\widetilde{\theta}}{16\pi^{2}}\int_{}^{}\widetilde{f} \star \widetilde{f} \\
&= \frac{3\theta_{2}}{16\pi^{2}}\int_{}^{}{tr\left( \left\lbrack g_{2} \otimes \mathbf{1}_{3} \right\rbrack \star \left\lbrack g_{2} \otimes \mathbf{1}_{3} \right\rbrack \right)} + \frac{2\theta_{3}}{16\pi^{2}}\int_{}^{}{tr\left( \left\lbrack g_{3} \otimes \mathbf{1}_{2} \right\rbrack \star \left\lbrack g_{3} \otimes \mathbf{1}_{2} \right\rbrack \right)}\\ 
&\ \ \ \ + \frac{\widetilde{\theta} - 18\theta_{2} - 12\theta_{3}}{16\pi^{2}36}\int_{}^{}{tr\left( \widetilde{f}\mathbf{1}_{6} \right)tr\left( \star \widetilde{f}\mathbf{1}_{6} \right)}.
\end{split}
\end{equation}
It can be seen that the ranges of \(\theta_{2}\), \(\theta_{3}\) and \(\widetilde{\theta}\) are \(\theta_{2},\theta_{3} \in \lbrack 0,2\pi)\) and \(\widetilde{\theta} \in \lbrack 0,72\pi)\).

Next, we will discuss the values of the \(\theta\) angles that are invariant under time reversal.

(1).\(G = \frac{\left\lbrack SU(3) \times SU(2)_{L} \times U(1)_{B - L} \right\rbrack \times SU(2)_{R}}{\Gamma} = \frac{\left\lbrack SU(3) \times SU(2)_{L} \times U(1)_{B - L} \right\rbrack}{\Gamma_{1}} \times \frac{SU(2)_{R}}{\Gamma_{2}}\)

Under this combination of the symmetry groups, the values of \(\Gamma_{1}\) and \(\Gamma_{2}\) can be as follows:

When \(\Gamma_{1} = 1,\Gamma_{2} = 1\), the range of each \(\theta\) angle is \(\theta_{2L},\theta_{2R},\theta_{3},\widetilde{\theta} \in \lbrack 0,2\pi)\). The time-reversal invariant values of the \(\theta\) angles are \(\theta_{2L},\theta_{2R},\theta_{3},\widetilde{\theta} = 0,\pi\).

When \(\Gamma_{1} = Z_{3},\Gamma_{2} = 1\), the range of each \(\theta\) angle is \(\theta_{2L},\theta_{2R},\theta_{3} \in \lbrack 0,2\pi);\widetilde{\theta} \in \lbrack 0,18\pi)\). The time-reversal invariant values of the \(\theta\) angles are \(\theta_{2L},\theta_{2R} = 0,\pi;\left\{ \begin{matrix} \theta_{3} = 0 \rightarrow \widetilde{\theta} = 0,9\pi \\ \theta_{3} = \pi \rightarrow \widetilde{\theta} = 3\pi,12\pi \\ \end{matrix} \right.\ \) .

When \(\Gamma_{1} = Z_{2L},\Gamma_{2} = 1\), the range of each \(\theta\) angle is \(\theta_{2L},\theta_{2R},\theta_{3} \in \lbrack 0,2\pi);\widetilde{\theta} \in \lbrack 0,8\pi)\). The time-reversal invariant values of the \(\theta\) angles are \(\theta_{2R},\theta_{3} = 0,\pi;\left\{ \begin{matrix} \theta_{2L} = 0 \rightarrow \widetilde{\theta} = 0,4\pi \\ \theta_{2L} = \pi \rightarrow \widetilde{\theta} = 2\pi,6\pi \\ \end{matrix} \right.\ \).

When \(\Gamma_{1} = Z_{3} \times Z_{2L} = Z_{6L},\Gamma_{2} = 1\), the range of each \(\theta\) angle is \(\theta_{2L},\theta_{2R},\theta_{3} \in \lbrack 0,2\pi);\widetilde{\theta} \in \lbrack 0,72\pi)\). The time-reversal invariant values of the \(\theta\) angles are \(\theta_{2R} = 0,\pi;\left\{ \begin{matrix} \theta_{3} = 0\left\{ \begin{matrix} \theta_{2L} = 0 \rightarrow \widetilde{\theta} = 0,36\pi \\ \theta_{2L} = \pi \rightarrow \widetilde{\theta} = 18\pi,54\pi \\ \end{matrix} \right.\  \\ \theta_{3} = \pi\left\{ \begin{matrix} \theta_{2L} = 0 \rightarrow \widetilde{\theta} = 12\pi,48\pi \\ \theta_{2L} = \pi \rightarrow \widetilde{\theta} = 30\pi,66\pi \\ \end{matrix} \right.\  \\ \end{matrix} \right.\ \).

The value of \(\Gamma_{2}\) is independent of \(\Gamma_{1}\). When \(\Gamma_{2} = Z_{2R}\), the range of \(\theta_{2R}\) in the above cases changes from \(\lbrack 0,2\pi)\) to \(\lbrack 0,4\pi)\). The time-reversal invariant values of the \(\theta_{2R}\) angle become \(\theta_{2R} = 0,2\pi\).

(2).\(G = \frac{\left\lbrack SU(3) \times SU(2)_{L} \right\rbrack \times \left\lbrack U(1)_{B - L} \times SU(2)_{R} \right\rbrack}{\Gamma} = \frac{\left\lbrack U(1)_{B - L} \times SU(2)_{R} \right\rbrack}{\Gamma_{1}} \times \frac{\left\lbrack SU(3) \times SU(2)_{L} \right\rbrack}{\Gamma_{2}}\)

Under this combination of the symmetry groups, the values of \(\Gamma_{1}\) and \(\Gamma_{2}\) can be as follows:

When \(\Gamma_{1} = 1,\Gamma_{2} = 1\), the range of each \(\theta\) angle is \(\theta_{2L},\theta_{2R},\theta_{3},\widetilde{\theta} \in \lbrack 0,2\pi)\). The time-reversal invariant values of the \(\theta\) angles are \(\theta_{2L},\theta_{2R},\theta_{3},\widetilde{\theta} = 0,\pi\).

When \(\Gamma_{1} = Z_{2R},\Gamma_{2} = 1\), the range of each \(\theta\) angle is \(\theta_{2L},\theta_{2R},\theta_{3} \in \lbrack 0,2\pi);\widetilde{\theta} \in \lbrack 0,8\pi)\). The time-reversal invariant values of the \(\theta\) angles are \(\theta_{2L},\theta_{3} = 0,\pi;\left\{ \begin{matrix} \theta_{2R} = 0 \rightarrow \widetilde{\theta} = 0,4\pi \\ \theta_{2R} = \pi \rightarrow \widetilde{\theta} = 2\pi,6\pi \\ \end{matrix} \right.\ \).

The value of \(\Gamma_{2}\) is independent of \(\Gamma_{1}\).

When \(\Gamma_{2} = Z_{2L}\), the range of \(\theta_{2L}\) in the above cases changes from \(\lbrack 0,2\pi)\) to \(\lbrack 0,4\pi)\). The time-reversal invariant values of the \(\theta_{2L}\) angle change to \(\theta_{2L} = 0,2\pi\).

When \(\Gamma_{2} = Z_{3}\), the range of \(\theta_{3}\) in the above cases changes from \(\lbrack 0,2\pi)\) to \(\lbrack 0,6\pi)\). The time-reversal invariant values of the \(\theta_{3}\) angle change to \(\theta_{3} = 0,3\pi\).

When \(\Gamma_{2} = Z_{2L} \times Z_{3} = Z_{6L}\), the range of \(\theta_{2L}\) in the above cases changes from \(\lbrack 0,2\pi)\) to \(\lbrack 0,4\pi)\), the range of \(\theta_{3}\) in the above cases changes from \(\lbrack 0,2\pi)\) to \(\lbrack 0,6\pi)\). The time-reversal invariant values of the \(\theta_{2L}\) angle change to \(\theta_{2L} = 0,2\pi\), the time-reversal invariant values of the \(\theta_{3}\) angle changes to \(\theta_{3} = 0,3\pi\).

(3).\(G = \frac{\left\lbrack SU(3) \times SU(2)_{L} \right\rbrack \times \left\lbrack SU(2)_{R} \right\rbrack \times \left\lbrack U(1)_{B - L} \right\rbrack}{\Gamma} = \frac{\left\lbrack SU(3) \times SU(2)_{L} \right\rbrack}{\Gamma_{1}} \times \frac{SU(2)_{R}}{\Gamma_{2}} \times \left\lbrack U(1)_{B - L} \right\rbrack\)

Under this combination of the symmetry groups, the values of \(\Gamma_{1}\) and \(\Gamma_{2}\) can be as follows:

When \(\Gamma_{1} = 1,\Gamma_{2} = 1\), the range of each \(\theta\) angle is \(\theta_{2L},\theta_{2R},\theta_{3},\widetilde{\theta} \in \lbrack 0,2\pi)\). The time-reversal invariant values of the \(\theta\) angles are \(\theta_{2L},\theta_{2R},\theta_{3},\widetilde{\theta} = 0,\pi\).

When \(\Gamma_{1} = Z_{2L},\Gamma_{2} = 1\), the range of each \(\theta\) angle is \(\theta_{2L} \in \lbrack 0,4\pi);\theta_{2R},\theta_{3},\widetilde{\theta} \in \lbrack 0,2\pi)\). The time-reversal invariant values of the \(\theta\) angles are \(\theta_{2L} = 0,2\pi;\theta_{2R},\theta_{3},\widetilde{\theta} = 0\).

When \(\Gamma_{1} = Z_{3},\Gamma_{2} = 1\), the range of each \(\theta\) angle is \(\theta_{2L},\theta_{2R},\widetilde{\theta} \in \lbrack 0,2\pi);\theta_{3} \in \lbrack 0,6\pi)\). The time-reversal invariant values of the \(\theta\) angles are \(\theta_{2L},\theta_{2R},\widetilde{\theta} = 0;\theta_{3} = 0,3\pi\).

When \(\Gamma_{1} = Z_{3} \times Z_{2L} = Z_{6L},\Gamma_{2} = 1\), the range of each \(\theta\) angle is \(\theta_{2L} \in \lbrack 0,4\pi);\theta_{2R},\widetilde{\theta} \in \lbrack 0,2\pi);\theta_{3} \in \lbrack 0,6\pi)\). The time-reversal invariant values of the \(\theta\) angles are \(\theta_{2L} = 0,2\pi;\theta_{2R},\widetilde{\theta} = 0,\pi;\theta_{3} = 0,3\pi\).

The values of \(\Gamma_{2}\) are independent of \(\Gamma_{1}\).

When \(\Gamma_{2} = Z_{2R}\), the range of \(\theta_{2R}\) in the above cases changes from \(\lbrack 0,2\pi)\) to \(\lbrack 0,4\pi)\). The time-reversal invariant values of the \(\theta_{2R}\) angle change to \(\theta_{2R} = 0,2\pi\).

(4).\(G = \frac{\left\lbrack SU(3) \times U(1)_{B - L} \right\rbrack \times SU(2)_{L} \times SU(2)_{R}}{\Gamma} = \frac{\left\lbrack SU(3) \times U(1)_{B - L} \right\rbrack}{\Gamma_{1}} \times \frac{SU(2)_{R}}{\Gamma_{2}} \times \frac{SU(2)_{L}}{\Gamma_{3}}\)

Under this combination of the symmetry groups, the values of \(\Gamma_{1},\Gamma_{2}\) and \(\Gamma_{3}\) can be as follows:

When \(\Gamma_{1} = 1,\Gamma_{2} = 1,\Gamma_{3} = 1\), the range of each \(\theta\) angle is \(\theta_{2L},\theta_{2R},\theta_{3},\widetilde{\theta} \in \lbrack 0,2\pi)\). The time-reversal invariant values of the \(\theta\) angles are \(\theta_{2L},\theta_{2R},\theta_{3},\widetilde{\theta} = 0,\pi\).

When \(\Gamma_{1} = Z_{3},\Gamma_{2} = 1,\Gamma_{3} = 1\), the range of each \(\theta\) angle is \(\theta_{2L},\theta_{2R},\theta_{3} \in \lbrack 0,2\pi);\widetilde{\theta} \in \lbrack 0,18\pi)\). The time-reversal invariant values of the \(\theta\) angles are \(\theta_{2R},\theta_{2L} = 0,\pi;\left\{ \begin{matrix} \theta_{3} = 0 \rightarrow \widetilde{\theta} = 0,9\pi \\ \theta_{3} = \pi \rightarrow \widetilde{\theta} = 3\pi,12\pi \\ \end{matrix} \right.\ \).

The values of \(\Gamma_{2},\Gamma_{3}\) are independent of \(\Gamma_{1}\).

When \(\Gamma_{2} = Z_{2R}\), the range of \(\theta_{2R}\) in the above cases changes from \(\lbrack 0,2\pi)\) to \(\lbrack 0,4\pi)\). The time-reversal invariant values of the \(\theta_{2R}\) angle change to \(\theta_{2R} = 0,2\pi\). When \(\Gamma_{3} = Z_{2L}\), the range of \(\theta_{2L}\) in the above cases changes from \(\lbrack 0,2\pi)\) to \(\lbrack 0,4\pi)\). The time-reversal invariant values of the \(\theta_{2L}\) angle change to \(\theta_{2L} = 0,2\pi\).

(5).\(G = \frac{SU(3) \times \left\lbrack SU(2)_{L} \times U(1)_{B - L} \right\rbrack \times SU(2)_{R}}{\Gamma} = \frac{\left\lbrack SU(2)_{L} \times U(1)_{B - L} \right\rbrack}{\Gamma_{1}} \times \frac{SU(2)_{R}}{\Gamma_{2}} \times \frac{SU(3)}{\Gamma_{3}}\)

Under this combination of the symmetry groups, the values of \(\Gamma_{1},\Gamma_{2}\) and \(\Gamma_{3}\) can be as follows:

When \(\Gamma_{1} = 1,\Gamma_{2} = 1,\Gamma_{3} = 1\), the range of each \(\theta\) angle is \(\theta_{2L},\theta_{2R},\theta_{3},\widetilde{\theta} \in \lbrack 0,2\pi)\). The time-reversal invariant values of the \(\theta\) angles are \(\theta_{2L},\theta_{2R},\theta_{3},\widetilde{\theta} = 0,\pi\).

When \(\Gamma_{1} = Z_{2L},\Gamma_{2} = 1,\Gamma_{3} = 1\), the range of each \(\theta\) angle is \(\theta_{2L},\theta_{2R},\theta_{3} \in \lbrack 0,2\pi);\widetilde{\theta} \in \lbrack 0,8\pi)\). The time-reversal invariant values of the \(\theta\) angles are \(\theta_{2R},\theta_{3} = 0,\pi;\left\{ \begin{matrix} \theta_{2L} = 0 \rightarrow \widetilde{\theta} = 0,4\pi \\ \theta_{2L} = \pi \rightarrow \widetilde{\theta} = 2\pi,6\pi \\ \end{matrix} \right.\ \).

The values of \(\Gamma_{2},\Gamma_{3}\) are independent of \(\Gamma_{1}\).

When \(\Gamma_{2} = Z_{2R}\), the range of \(\theta_{2R}\) in the above cases changes from \(\lbrack 0,2\pi)\) to \(\lbrack 0,4\pi)\). The time-reversal invariant values of the \(\theta_{2R}\) angle change to \(\theta_{2R} = 0,2\pi\). When \(\Gamma_{3} = Z_{3}\), the range of \(\theta_{3}\) in the above cases changes from \(\lbrack 0,2\pi)\) to \(\lbrack 0,6\pi)\). The time-reversal invariant values of the \(\theta_{3}\) angle change to \(\theta_{3} = 0,3\pi\).

(6).\(G = \frac{\left\lbrack SU(3) \right\rbrack \times \left\lbrack SU(2)_{L} \right\rbrack \times \left\lbrack SU(2)_{R} \right\rbrack \times \lbrack U(1)_{B - L}\rbrack}{\Gamma} = \frac{SU(2)_{L}}{\Gamma_{1}} \times \frac{SU(2)_{R}}{\Gamma_{2}} \times \frac{SU(3)}{\Gamma_{3}} \times U(1)_{B - L}\)

Under this combination of the symmetry groups, the values of \(\Gamma_{1},\Gamma_{2}\) and \(\Gamma_{3}\) can be as follows:

When \(\Gamma_{1} = 1,\Gamma_{2} = 1,\Gamma_{3} = 1\), the range of each \(\theta\) angle is \(\theta_{2L},\theta_{2R},\theta_{3},\widetilde{\theta} \in \lbrack 0,2\pi)\). The time-reversal invariant values of the \(\theta\) angles are \(\theta_{2L},\theta_{2R},\theta_{3},\widetilde{\theta} = 0,\pi\).

The values of \(\Gamma_{1},\Gamma_{2},\Gamma_{3}\) is independent of each other. When \(\Gamma_{1} = Z_{2L}\), the range of \(\theta_{2L}\) in the above cases changes from \(\lbrack 0,2\pi)\) to \(\lbrack 0,4\pi)\). The time-reversal invariant values of the \(\theta_{2L}\) angle change to \(\theta_{2L} = 0,2\pi\). When \(\Gamma_{2} = Z_{2R}\), the range of \(\theta_{2R}\) in the above cases changes from \(\lbrack 0,2\pi)\) to \(\lbrack 0,4\pi)\). The time-reversal invariant values of the \(\theta_{2R}\) angle change to \(\theta_{2R} = 0,2\pi\). When \(\Gamma_{3} = Z_{3}\), the range of \(\theta_{3}\) in the above cases changes from \(\lbrack 0,2\pi)\) to \(\lbrack 0,6\pi)\). The time-reversal invariant values of the \(\theta_{3}\) angle change to \(\theta_{3} = 0,3\pi\).

Due to the \(SU(2)_{L}\) and \(SU(2)_{R}\) symmetry in the left-right symmetric model, some results were omitted above. These results can be obtained by replacing \(SU(2)_{L}\) with \(SU(2)_{R}\) and \(SU(2)_{R}\) with \(SU(2)_{L}\) in the above text.

\section{Electroweak symmetry breaking}

Next, we will discuss the effect of electroweak symmetry breaking on the spectrum of line operators. As is well known, the process of electroweak breaking in the Standard Model involves the breaking of the \(SU(2) \times U(1)_{Y}\) part of the \(SU(3) \times SU(2) \times U(1)_{Y}\) to the electromagnetic symmetry \(U(1)_{em}\). In the left-right symmetric model, the breaking occurs in two steps: first, the \(SU(2)_{R} \times U(1)_{B - L}\) part of the \(SU(3) \times SU(2)_{L} \times SU(2)_{R} \times U(1)_{B - L}\) breaks to \(U(1)_{Y}\), reducing the left-right symmetric gauge group to the Standard Model gauge group \(SU(3) \times SU(2) \times U(1)_{Y}\), and in the second step, the Standard Model gauge group breaks to \(SU(3) \times U(1)_{em}\).

The Gell-Mann--Nishijima relation in the left-right symmetric model is as follows:
\begin{equation}
Q_{em} = I_{3L} + I_{3R} + \frac{B - L}{2}.
\end{equation}
where \(Q_{em}\) is the charge of the electromagnetic \(U(1)_{em}\), \(I_{3L}\) and \(I_{3R}\) are the third components of the \(SU(2)_{L}\) and \(SU(2)_{R}\) generators, respectively\cite{mohapatra1980local}. They are related to the charges of the corresponding groups as follows: \(I_{3L} = \frac{1}{2}z_{2L}^{e}\) and \(I_{3R} = \frac{1}{2}z_{2R}^{e}\). Comparing this with the Gell-Mann--Nishijima relation in the Standard Model:
\begin{equation}
Q_{em} = I_{3L} + \frac{q_{Y}}{6},
\end{equation}
we get:
\begin{equation}
\frac{q_{Y}}{6} = I_{3R} + \frac{B - L}{2} = \frac{1}{2}z_{2R}^{e} + \frac{q}{6}.
\end{equation}

In the left-right symmetric model, the breaking of \(SU(2)_{R} \times U(1)_{B - L}\) to \(U(1)_{Y}\) is achieved through the non-zero vacuum expectation value of the Higgs triplet \(\Delta_{R} \in \left( \mathbf{1}_{L}\mathbf{,}\mathbf{3}_{R}\mathbf{,}2 \right)\). Substituting the charges of the Higgs triplet \(\Delta_{R}\) into the quantization condition, we get:
\begin{equation}
z_{2R}^{m} = 6g + n.
\end{equation}
Substituting this back into the quantization condition \(\left( 3z_{2R}^{e}z_{2R}^{m} - 6qg \right) = 6n\), we get:
\begin{equation}
\left( 18gz_{2R}^{e} + 3mz_{2R}^{e} - 6qg \right) = 6n.
\end{equation}
This can be rewritten as:
\begin{equation}
6g\left( \frac{1}{2}z_{2R}^{e} - \frac{1}{6}q \right) = n.
\end{equation}
Noting that \(\frac{q_{Y}}{6} = \frac{1}{2}z_{2R}^{e} + \frac{q}{6}\), this can be written as:
\begin{equation}
gq_{Y} = n.
\end{equation}
Thus, the magnetic charge of \(U(1)_{Y}\) is \(g_{Y} = g\). Therefore, we can write the spectrum of line operators after the breaking of \(SU(2)_{R} \times U(1)_{B - L}\) to \(U(1)_{Y}\).

(1).\(\widetilde{G} = (SU(3) \times SU(2)_{L} \times U(1)_{B - L}) \times \left( SU(2)_{R} \right)\)

In this gauge group combination, the choice of \(\Gamma\) affects the spectrum of line operators.

When \(\Gamma = 1,Z_{2R}\), the electric charges of the Wilson line can take the following values: \(z_{2L}^{e} = 0,1;z_{3}^{e} = 0,1,2;q_{Y} \in Z\). The magnetic charges of the \textquotesingle t Hooft line can take the values \(z_{2L}^{m},z_{3}^{m} = 0;g_{Y} \in Z\).

When \(\Gamma = Z_{2L},Z_{2L} \times Z_{2R}\), the electric charges of the Wilson line can take the following values: \(z_{2L}^{e} = 0;z_{3}^{e} = 0,1,2;q_{Y} \in Z\). The magnetic charges of the \textquotesingle t Hooft line can take the values \(z_{2L}^{m} = 0,1;z_{3}^{m} = 0;g_{Y} \in Z\).

When \(\Gamma = Z_{3},Z_{2R} \times Z_{3}\), the electric charges of the Wilson line can take the following values: \(z_{2L}^{e} = 0,1;z_{3}^{e} = 0,1,2;q_{Y} = 3n - z_{3}^{e}\). The magnetic charges of the \textquotesingle t Hooft line can take the values \(z_{2L}^{m} = 0;z_{3}^{m} = 0,1,2;g_{Y} = m - \frac{1}{3}z_{3}^{m}\).

When \(\Gamma = Z_{2L} \times Z_{3},Z_{2L} \times Z_{2R} \times Z_{3}\), the electric charges of the Wilson line can take the following values: \(z_{2L}^{e} = 0;z_{3}^{e} = 0,1,2;q_{Y} = 3n - z_{3}^{e}\). The magnetic charges of the \textquotesingle t Hooft line can take the values \(z_{2L}^{m} = 0,1;z_{3}^{m} = 0,1,2;g_{Y} = m - \frac{1}{3}z_{3}^{m}\).

(2).\(\widetilde{G} = \left\lbrack SU(3) \times SU(2)_{L} \right\rbrack \times \left\lbrack SU(2)_{R} \times U(1)_{B - L} \right\rbrack\)

When \(\Gamma = 1,Z_{2R}\), the electric charges of the Wilson line can take the following values: \(z_{2L}^{e} = 0,1;z_{3}^{e} = 0,1,2;q_{Y} = 3n\). The magnetic charges of the \textquotesingle t Hooft line can take the values \(z_{2L}^{m} = 0;z_{3}^{m} = 0;g_{Y} = m\).

When \(\Gamma = Z_{2L},Z_{2L} \times Z_{2R}\), the electric charges of the Wilson line can take the following values: \(z_{2L}^{e} = 0;z_{3}^{e} = 0,1,2;q_{Y} = 3n\). The magnetic charges of the \textquotesingle t Hooft line can take the values \(z_{2L}^{m} = 0,1;z_{3}^{m} = 0;g_{Y} = m\).

When \(\Gamma = Z_{3},Z_{2R} \times Z_{3}\), the electric charges of the Wilson line can take the following values: \(z_{2L}^{e} = 0,1;z_{3}^{e} = 0;q_{Y} = 3n\). The magnetic charges of the \textquotesingle t Hooft line can take the values \(z_{2L}^{m} = 0;z_{3}^{m} = 0,1,2;g_{Y} = m\).

When \(\Gamma = Z_{2L} \times Z_{3},Z_{2L} \times Z_{2R} \times Z_{3}\), the electric charges of the Wilson line can take the following values: \(z_{2L}^{e} = 0;z_{3}^{e} = 0;q_{Y} = 3n\). The magnetic charges of the \textquotesingle t Hooft line can take the values \(z_{2L}^{m} = 0,1;z_{3}^{m} = 0,1,2;g_{Y} = m\).

For $\widetilde{G} = \left\lbrack SU(3) \times SU(2)_{R} \times U(1)_{B - L} \right\rbrack \times SU(2)_{L},  \left\lbrack SU(3) \times U(1)_{B - L} \right\rbrack \times SU(2)_{L} \times SU(2)_{R}$ cases, the spectrum of line operators is the same as the (1) case.

For $\widetilde{G} = \left\lbrack SU(3) \times SU(2)_{R} \right\rbrack \times \left\lbrack SU(2)_{L} \times U(1)_{B - L} \right\rbrack , \left\lbrack SU(3) \times SU(2)_{L} \right\rbrack \times \left\lbrack SU(2)_{R} \right\rbrack \times \left\lbrack U(1)_{B - L} \right\rbrack , \left\lbrack SU(3) \times SU(2)_{R} \right\rbrack \times \left\lbrack SU(2)_{L} \right\rbrack \times \left\lbrack U(1)_{B - L} \right\rbrack , SU(3) \times SU(2)_{L} \times SU(2)_{R} \times U(1)_{B - L}, SU(3) \times \left\lbrack SU(2)_{L} \times U(1)_{B - L} \right\rbrack \times SU(2)_{R}, SU(3) \times \left\lbrack SU(2)_{R} \times U(1)_{B - L} \right\rbrack \times SU(2)_{L}$ cases, the spectrum of line operators is the same as the (2) case.

Next, let's discuss the second step of the left-right symmetric model breaking to electromagnetism, which is the process of breaking from the Standard Model's \(SU(3) \times SU(2)_{L} \times U(1)_{Y}\) to \(SU(3) \times U(1)_{em}\).

The Gell-Mann--Nishijima relation for the breaking of the Standard Model to the electromagnetic theory is:
\begin{equation}
Q_{em} = I_{3L} + \frac{q_{Y}}{6} = \frac{1}{2}z_{2L}^{e} + \frac{q_{Y}}{6}.
\end{equation}
The process of breaking the Standard Model to the electromagnetic theory is completed by the Higgs triplet \(\left( \mathbf{2}_{SU(2)_{L}\ }\mathbf{,}\mathbf{1}_{SU(3)}\mathbf{,}3_{U(1)_{Y}} \right)\). Substituting the charges corresponding to each gauge group into the Dirac quantization condition, we get:
\begin{equation}
3z_{2L}^{m} - 3 \times 6g_{Y} = 6n \Longrightarrow z_{2L}^{m} = 6g_{Y} + 2n.
\end{equation}
Substituting this relation back into the quantization condition, we get:
\begin{equation}
\left( 18g_{Y}z_{2L}^{e} - 6q_{Y}g_{Y} \right) = 6n \Longrightarrow 6g_{Y}\left( \frac{1}{2}z_{2L}^{e} - \frac{1}{6}q_{Y} \right) = n.
\end{equation}
From the Gell-Mann--Nishijima relation \(Q_{em} = \frac{1}{2}z_{2L}^{e} + \frac{q_{Y}}{6}\), we get:
\begin{equation}
6g_{Y}Q_{em} = n.
\end{equation}
The corresponding magnetic charge \(G_{em}\) of the electromagnetic \(U(1)_{em}\) is:
\begin{equation}
G_{em} = 6g_{Y}.
\end{equation}

Thus, we can write the spectrum of the line operators for the left-right symmetric breaking to the electromagnetic model.

(1).\(\widetilde{G} = \left\lbrack SU(3) \times SU(2)_{L} \times U(1)_{B - L} \right\rbrack \times SU(2)_{R}\)

For \(\Gamma = 1,Z_{2L}\), the electric charges of the Wilson lines can take the following values: \(z_{3}^{e} = 0,1,2;Q_{em} = \frac{n}{6}\), and the corresponding magnetic charges of the \textquotesingle t Hooft line can take the values \(z_{3}^{m} = 0;G_{em} = 6m\).

For \(\Gamma = Z_{3},Z_{2L} \times Z_{3}\), the electric charges of the Wilson lines can take the following values: \(z_{3}^{e} = 0,1,2;\left\{ \begin{matrix} z_{3}^{e} = 0 \rightarrow Q_{em} = 0 \\ z_{3}^{e} = 1 \rightarrow Q_{em} = \frac{2}{6} \\ z_{3}^{e} = 2 \rightarrow Q_{em} = \frac{1}{6} \\ \end{matrix} \right.\ \), and the corresponding magnetic charges of the \textquotesingle t Hooft line can take the values \(z_{3}^{m} = 0,1,2;\left\{ \begin{matrix} z_{3}^{m} = 0 \rightarrow G_{em} = 0 \\ z_{3}^{m} = 1 \rightarrow G_{em} = 4 \\ z_{3}^{m} = 2 \rightarrow G_{em} = 2 \\ \end{matrix} \right.\ \).

(2).\(\widetilde{G} = \left\lbrack SU(3) \times SU(2)_{L} \right\rbrack \times \left\lbrack SU(2)_{R} \times U(1)_{B - L} \right\rbrack\)

For \(\Gamma = 1,Z_{2L}\), the electric charges of the Wilson lines can take the following values: \(z_{3}^{e} = 0,1,2;Q_{em} = \frac{n}{2}\), and the corresponding magnetic charges of the \textquotesingle t Hooft line can take the values \(z_{3}^{m} = 0;G_{em} = 6m\).

For \(\Gamma = Z_{3},Z_{2L} \times Z_{3}\), the electric charges of the Wilson lines can take the following values: \(z_{3}^{e} = 0;Q_{em} = \frac{n}{2}\), and the corresponding magnetic charges of the \textquotesingle t Hooft line can take the values \(z_{3}^{m} = 0,1,2;G_{em} = 6m\).

In $\widetilde{G} = \left\lbrack SU(3) \times SU(2)_{R} \times U(1)_{B - L} \right\rbrack \times SU(2)_{L},  \left\lbrack SU(3) \times U(1)_{B - L} \right\rbrack \times SU(2)_{L} \times SU(2)_{R}$ cases, the spectrum of line operators is the same as the (1) case.

In $\widetilde{G} = \left\lbrack SU(3) \times SU(2)_{R} \right\rbrack \times \left\lbrack SU(2)_{L} \times U(1)_{B - L} \right\rbrack ,  \left\lbrack SU(3) \times SU(2)_{L} \right\rbrack \times \left\lbrack SU(2)_{R} \right\rbrack \times \left\lbrack U(1)_{B - L} \right\rbrack ,  \left\lbrack SU(3) \times SU(2)_{R} \right\rbrack \times \left\lbrack SU(2)_{L} \right\rbrack \times \left\lbrack U(1)_{B - L} \right\rbrack , SU(3) \times SU(2)_{L} \times SU(2)_{R} \times U(1)_{B - L},  SU(3) \times \left\lbrack SU(2)_{L} \times U(1)_{B - L} \right\rbrack \times SU(2)_{R},  SU(3) \times \left\lbrack SU(2)_{R} \times U(1)_{B - L} \right\rbrack \times SU(2)_{L}$ cases, the spectrum of line operators is the same as the (2) case.

Next, we will discuss the differences of \(\theta\) angle between the process of left-right symmetry breaking to electromagnetic theory and the Standard Model. As mentioned above, the process of left-right symmetry breaking to electromagnetic theory is different from the one-step breaking to electromagnetism in the Standard Model. The left-right symmetry breaking process first involves the breaking of \(SU(2)_{R} \times U(1)_{B - L}\) to the Standard Model's \(U(1)_{Y}\), followed by a process similar to the Standard Model's breaking, i.e. \(SU(2)_{L} \times U(1)_{Y}\) breaking to the electromagnetic theory's \(U(1)_{em}\). This process also involves changes in the \(\theta\) terms of various gauge groups before and after the breaking.

First, we write the \(\theta\) terms of the left-right symmetric model\cite{choi2024quantization,reece2023axion}:
\begin{equation}
\begin{split}
S_{\theta} &= \frac{\theta_{3}g_{3}^{2}}{32\pi^{2}}\int_{}^{}{G_{\mu\nu}^{a} \star G^{a\mu\nu}\ } + \frac{\theta_{2L}g_{2L}^{2}}{32\pi^{2}}\int_{}^{}{\left( F_{L} \right)_{\mu\nu}^{a} \star \left( F_{L} \right)^{a\mu\nu}\ } \\
&+ \frac{\theta_{2R}g_{2R}^{2}}{32\pi^{2}}\int_{}^{}{\left( F_{R} \right)_{\mu\nu}^{a} \star \left( F_{R} \right)^{a\mu\nu}\ } + \frac{\theta_{1}g_{1}^{2}}{16\pi^{2}}\frac{1}{36}\int_{}^{}{B_{\mu\nu} \star B^{\mu\nu}\ },
\end{split}
\end{equation}
where ${G_{\mu\nu}^{a}}$, ${\left( F_{L} \right)_{\mu\nu}^{a}}$, ${\left( F_{R} \right)_{\mu\nu}^{a}}$, ${B_{\mu\nu}}$ are field strengthes corresponding to $SU(3)$, $SU(2)_{L}$, $SU(2)_{R}$, $U(1)_{B - L}$ gauge groups respectively. Next, we write down the relation between the field strengths of the gauge groups before and after the breaking of \(SU(2)_{R} \times U(1)_{B - L}\) to the Standard Model's \(U(1)_{Y}\), similar to the electroweak breaking process in the Standard Model:
\begin{equation}
\left( F_{R} \right)_{\mu\nu}^{3} = \frac{e_{Y}}{g_{2R}}\left( F_{Y} \right)_{\mu\nu} + \ldots,
\end{equation}
\begin{equation}
B_{\mu\nu} = \frac{e_{Y}}{g_{1}}\left( F_{Y} \right)_{\mu\nu} + \ldots.
\end{equation}
The above two equations omit terms unrelated to \(U(1)_{Y}\) after breaking. Substituting these into the complete \(\theta\) term, we get:
\begin{equation}
\begin{split}
&\frac{\theta_{2R}g_{2R}^{2}}{32\pi^{2}}\int_{}^{}{\left( F_{R} \right)_{\mu\nu}^{a} \star \left( F_{R} \right)^{a\mu\nu}\ } + \frac{\theta_{1}g_{1}^{2}}{16\pi^{2}}\frac{1}{36}\int_{}^{}{B_{\mu\nu} \star B^{\mu\nu}\ } \\
&= \frac{\theta_{2R}e_{Y}^{2}}{32\pi^{2}}\int_{}^{}{\left( F_{Y} \right)_{\mu\nu} \star \left( F_{Y} \right)^{\mu\nu}} + \frac{\theta_{1}e_{Y}^{2}}{16\pi^{2}}\frac{1}{36}\int_{}^{}{\left( F_{Y} \right)_{\mu\nu} \star \left( F_{Y} \right)^{\mu\nu}\ }\\
&= \left\lbrack \frac{\theta_{2R}}{2} + \frac{\theta_{1}}{36} \right\rbrack\frac{e_{Y}^{2}}{16\pi^{2}}\int_{}^{}{\left( F_{Y} \right)_{\mu\nu} \star \left( F_{Y} \right)^{\mu\nu}\ } = \left\lbrack \frac{18\theta_{2R} + \theta_{1}}{36} \right\rbrack\frac{e_{Y}^{2}}{16\pi^{2}}\int_{}^{}{\left( F_{Y} \right)_{\mu\nu} \star \left( F_{Y} \right)^{\mu\nu}\ }.
\end{split}
\end{equation}
It can be seen that the relation between \(\theta\) before and after breaking is:
\begin{equation}
\frac{\theta_{Y}}{36} = \left\lbrack \frac{18\theta_{2R} + \theta_{1}}{36} \right\rbrack.
\end{equation}

The periodicity of \(\theta_{Y}\) will depend on the choice of \(\Gamma\). When \(\Gamma = 1,Z_{2R}\), \(\theta_{Y} \in \left\lbrack 0,2\pi e_{Y}^{2} \right)\), where \(e_{Y}\) is the minimum electric charge of the \(U(1)_{Y}\) gauge group. When \(\Gamma = Z_{2L},Z_{2L} \times Z_{2R}\), \(\theta_{Y} \in \left\lbrack 0,8\pi e_{Y}^{2} \right)\). When \(\Gamma = Z_{3},Z_{3} \times Z_{2R}\), \(\theta_{Y} \in \left\lbrack 0,18\pi e_{Y}^{2} \right)\). When \(\Gamma = Z_{2L} \times Z_{3},Z_{2L} \times Z_{3} \times Z_{2R}\), \(\theta_{Y} \in \left\lbrack 0,72\pi e_{Y}^{2} \right)\). The \(\theta\) term after the first step of breaking is:
\begin{equation}
\begin{split}
S_{\theta} = \frac{\theta_{3}g_{3}^{2}}{32\pi^{2}}\int_{}^{}{G_{\mu\nu}^{a} \star G^{a\mu\nu}\ } + \frac{\theta_{2L}g_{2L}^{2}}{32\pi^{2}}\int_{}^{}{\left( F_{L} \right)_{\mu\nu}^{a} \star \left( F_{L} \right)^{a\mu\nu}\ } + \frac{\theta_{Y}}{36}\frac{e_{Y}^{2}}{16\pi^{2}}\int_{}^{}{\left( F_{Y} \right)_{\mu\nu} \star \left( F_{Y} \right)^{\mu\nu}\ }
\end{split}
\end{equation}
The relationship between the field strengths of the gauge fields before and after breaking to electromagnetism is as follows:
\begin{equation}
\left( F_{L} \right)_{\mu\nu}^{3} = \frac{e_{em}}{g_{2L}}\left( F_{em} \right)_{\mu\nu} + \ldots
\end{equation}
\begin{equation}
\left( F_{Y} \right)_{\mu\nu} = \frac{e_{em}}{e_{Y}}\left( F_{em} \right)_{\mu\nu} + \ldots
\end{equation}
Substituting these into the \(\theta\) term after the first step of breaking, we get:
\begin{equation}
\begin{split}
&\frac{\theta_{2L}g_{2L}^{2}}{32\pi^{2}}\int_{}^{}{\left( F_{L} \right)_{\mu\nu}^{a} \star \left( F_{L} \right)^{a\mu\nu}\ } + \frac{\theta_{Y}}{36}\frac{e_{Y}^{2}}{16\pi^{2}}\int_{}^{}{\left( F_{Y} \right)_{\mu\nu} \star \left( F_{Y} \right)^{\mu\nu}\ }\\ 
&= \frac{\theta_{2L}e_{em}^{2}}{32\pi^{2}}\int_{}^{}{\left( F_{em} \right)_{\mu\nu} \star \left( F_{em} \right)^{\mu\nu}\ } + \frac{\theta_{Y}}{36}\frac{e_{em}^{2}}{16\pi^{2}}\int_{}^{}{\left( F_{em} \right)_{\mu\nu} \star \left( F_{em} \right)^{\mu\nu}\ }\\
&= \left\lbrack \frac{18\theta_{2L} + \theta_{Y}}{36} \right\rbrack\frac{e_{em}^{2}}{16\pi^{2}}\int_{}^{}{\left( F_{em} \right)_{\mu\nu} \star \left( F_{em} \right)^{\mu\nu}\ }
\end{split}
\end{equation}
It can be seen that the relationship between \(\theta\) before and after breaking is:
\begin{equation}
\theta_{em} = \left\lbrack \frac{18\theta_{2L} + \theta_{Y}}{36} \right\rbrack
\end{equation}
Here, the periodicity of \(\theta_{em}\) will also depend on the choice of \(\Gamma\). When \(\Gamma = 1,Z_{2L}\), \(\theta_{em} \in \left\lbrack 0,2\pi e_{em}^{2} \right)\), where \(e_{Y}\) is the minimum electric charge of the \(U(1)_{Y}\) gauge group. When \(\Gamma = Z_{3},Z_{3} \times Z_{2L}\), \(\theta_{Y} \in \left\lbrack 0,18\pi e_{Y}^{2} \right)\).

\section{Summary}

This paper discussed issues related to line operators in left-right symmetric models. First, we derived the spectrum of line operators in all possible scenarios within left-right symmetric models. Due to differences in gauge group structures, left-right symmetric models have many more scenarios compared to the Standard Model, resulting in significant differences in the spectrum of line operators. We then discussed the range of the $\theta$ angle and the corresponding time-reversal invariant states for all choices of $\Gamma$. Similarly, due to the differences in gauge groups, the $\theta$ angle in left-right symmetric model also differs significantly from that in the Standard Model. Finally, we discussed the spectrum of line operators and the $\theta$ angle after the electroweak symmetry breaking. The results show that even when breaking down to electromagnetic theory, the resulting spectrum of line operators and $\theta$ angle still differ from those in the Standard Model. These differences provide potential for future experimental verification.

\section*{Acknowledgment}

This work is supported by the Natural Science Foundation of Chongqing, China ( Grant No. CSTB2022NSCQ-MSX0351 ).

\bibliography{ref}
\bibliographystyle{unsrt}

		\end{document}